**Communication**

# Programmable actuation of porous poly(ionic liquid) membranes by aligned carbon nanotubes

*Huijuan Lin,[1] Jiang Gong,[1] Michaela Eder,[2] Roman Schuetz,[2] Huisheng Peng,[3] John W. C. Dunlop,[2]\* and Jiayin Yuan[1]\**

[1]H. Lin, Dr. J. Gong, Dr. J. Yuan
Department of Colloid Chemistry, Max Planck Institute of Colloids and Interfaces
Am Mühlenberg 1 OT Golm, D-14476 Potsdam, Germany
E-Mail: jiayin.yuan@mpikg.mpg.de
[2]Dr. M. Eder, Dr. R. Schuetz, Dr. J. Dunlop
Department of Biomaterials, Max Planck Institute of Colloids and Interfaces
Am Mühlenberg 1 OT Golm, D-14476 Potsdam, Germany
E-Mail: john.dunlop@mpikg.mpg.de
[3]Prof. H. Peng
Department of Macromolecular Science and Laboratory of Advanced Materials, Fudan University
Shanghai 200438, China

Increasing interest has been recently directed to responsive materials towards external stimuli, *e.g.* pH in solution, humidity, solvent quality, ionic strength, light and temperature.[1-13] Such materials, when undergoing macroscopically detectable and preferentially reversible shape deformation, can present promising applications as sensors, motors, artificial muscles and electronic devices.[14-21] Besides improvement in material processing, a great deal of effort has been so far devoted to exploiting new chemical structures of building blocks and synthetic routes to develop novel responsive materials and mechanisms. In order to create an actuator from a responsive material that moves in a more complex manner than simply changing its volume upon stimulation, it is necessary to control where these volume changes occur. This can be realized by physically constraining the responsive material so that actuation only occurs in a particular direction (or in a given region of the material), which can be done by combining this responsive material with less-responsive or passive ones[22,23] or by controlling the spatial distribution of the stimulating signal (moisture, temperature etc.).[24,25] In one example of physical constraints, it has been shown that by reducing a graphene oxide fiber to hydrophobic graphene on only one side (e.g. the top half), a sensor could be made that curled upon changes in humidity; as a result of different response of the top (hydrophobic graphene) and bottom (hydrophilic graphene oxide) faces of the fiber towards moisture, the top face bent inwards with increasing humidity.[26] Other examples of using geometric constraints to control actuation include the classical bi-metallic strip,[27] complex swellable honeycombs,[28] and anisotropically patterned hydrogels.[23] Generally speaking, controlled actuation in terms of the shape and direction is of crucial importance, since it is required in the practical usage to fulfill a task-specific function. One promising area of inspiration for actuator design can be found in the plant world, as plant tissues are known to perform sophisticated shape changes in response to moisture.[29,30] The swelling of the hygroscopic polysaccharide matrix is controlled by the orientation of the anisotropic structural units of cellulose nano/microfibrils inside the cell wall. The orientation of these anisotropic building-blocks controls the swelling/shrinkage degree in different directions, thus a control of their orientation allows for only certain actuations upon external stimuli.[29-32]

Our group has previously reported porous polymer membrane actuators with ultrafast actuation kinetics and large scale locomotion towards an organic vapor phase.[33] The porous membrane was built up *via* electrostatic complexation between a hydrophobic cationic poly(ionic liquid) (PIL) and an *in-situ* neutralized multi-acid compound. PIL, a polymerization product of an ionic liquid,[34] was employed here as a type of innovative polyelectrolyte that facilitates the pore generation on account of their ionic, water-insoluble feature.[35] The role of the multi-acid was to balance the polycation charge, upon neutralization by ammonia, and meanwhile to electrostatically cross-link the PIL network to lock the porous state. The as-synthesized, dry porous membrane underwent actuation due to a gradient in the degree of electrostatic complexation (DEC, defined as the fraction of repeating units that undergo complexation with regard to the total repeating units in the PIL chain) between the PIL and the multi-acid, along the membrane cross-section, introduced by ammonia diffusion from top to bottom.[33] The inward bending of our polymer membrane actuator is determined by the chemical and structural gradient, that is, a higher crosslinking density on the top (stiff and with low swellability) decreases smoothly to the bottom (soft and highly swellable). The direction of maximum curvature however is difficult to control being dependent on the shape of the external boundary.[24] This is unfavorable in a practical use, as the application of actuators relies much on an accurate, predetermined movement.[36, 37] Our interest was triggered to introduce new structural elements to achieve a high level of control to dictate the actuation behavior.

Inspired by the function of cellulose fibers in plants, the incorporation of 1D structural units into such membranes offers a logical, promising approach and to test this idea we choose one of the most studied 1D nanomaterials, carbon nanotubes (CNTs). Aligned CNTs have been previously reported to interact with a conjugated polymer matrix to shape the chain conformation, which could be responsible for the observed anisotropic mechanical actuation behaviour of the composite.[5] Alternatively the presence of aligned stiff CNTs could simply constrain swelling of the surround polymer giving rise to anisotropic actuation. Additionally, due to the electronic structure of graphitic layers, their mechanical strength and flexibility,[38,39] the introduction of aligned CNTs into a polymer matrix will add extra merits into the actuator, *e.g.* high tensile strength along the fiber direction and electric conductivity. However, to structurally incorporate aligned CNTs into porous polymer membranes without interfering the pore structure is a challenging task itself. In this contribution, we demonstrate how to assemble aligned multi-walled CNTs into porous PIL membranes, and *via* this assembly how to precisely design and dictate the bending event of the membrane actuator as well as its electrical conductivity. This achievement further allows us to manipulate actuation behaviors at a high level of complexity and function.

The synthesis of CNT-doped porous polymer composite membrane is illustrated in **Figure 1**a. At first, poly(acrylic acid) (PAA) and a cationic PIL were dissolved in DMF. Since PAA in DMF is protonated thus non-charged, it is homogeneously mixed with PIL together. The cationic PIL used here, poly[3-cyanomethyl-1-vinylimidazolium bis(trifluoromethane sulfonyl)imide] (PIL) is synthesized through conventional radical polymerization of an ionic liquid monomer (3-cyanomethyl-1-vinylimidazolium bromide), followed by anion exchange with a lithium salt, lithium bis(trifluoromethane sulfonyl)imide (LiTf$_2$N) in aqueous solution to replace Br$^-$ by Tf$_2$N$^-$ (synthetic route towards the PIL shown in Figure S1, Supporting Information). The anion exchange produced the targeted hydrophobic PIL. The chemical structure of PIL was confirmed by $^1$H-NMR spectrum (Figure S2, Supporting Information), which matches well with that in literature.[40] To prepare the composite porous membrane, the aligned CNT sheets were dry-drawn from spinnable CNT arrays (Figure S3, Supporting Information) that were produced through chemical vapor deposition.[41] They were firstly paved onto a clean glass slide to form a flat sheet substrate. Figure 1b and 1c show typical scanning electron microscopy (SEM) images of bare aligned CNT sheets. Clearly, the CNTs stay highly oriented with open interstices, making it possible to infiltrate the polymer solution into the voids to prepare composite materials. The diameter of these CNTs is typically about 15 nm, one of which is visualized in a TEM image in Figure S4 (Supporting Information).

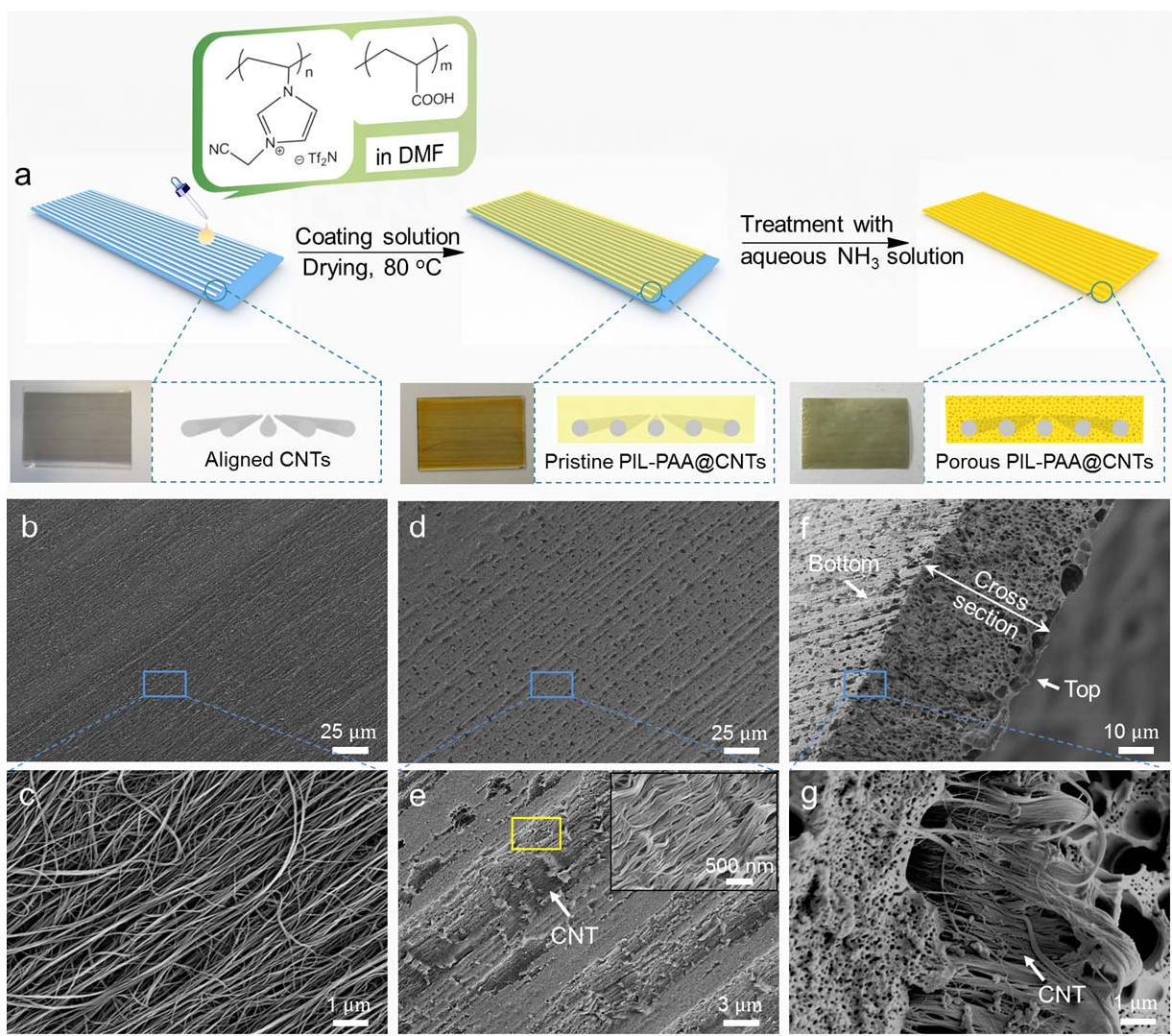

**Figure 1.** a) Schematic illustration of the fabrication of the composite porous membrane PIL-PAA@CNTs. b,c) SEM images of bare aligned CNT sheets in low and high magnifications. d,e) SEM images of PIL-PAA@CNTs from bottom view in low and high magnifications. Inset is the SEM image of the region marked with a yellow box in (e). f,g) SEM images of PIL-PAA@CNTs from side view in low and high magnifications, respectively.

The composite membrane is produced by coating a solution of PIL and PAA in DMF onto a pre-aligned CNT sheet on a glass plate. During this process, the interstitial voids among CNTs are occupied by the solution. The composite was then dried at 80 °C for 3 h resulting in a solid, transparent black film sticking to the glass surface. Here, the surface that sticks to the glass plate is termed as the BOTTOM surface, while the open side towards air as the TOP surface. The entire glass plate was then immersed in an aqueous ammonia solution (0.2 wt. %) for 2 h to promote the formation of a porous structure. During this process, ammonia diffused into the film from the top to the bottom, neutralized PAA into anionic poly(ammonium acrylate) and introduced *in-situ* inter-polyelectrolyte complexation between the PIL and PAA to produce the porous membrane. After this ammonia activation step, the membrane was easily peeled off. Figure 1d and 1e show SEM images of the as-synthesized PIL-PAA@CNTs composite membrane from a bottom view, where a flat surface is observed with the micrometer-sized carbon yarns occasionally visible as dark strips (indicated by white arrows). Comparison between Figure 1b and 1d indicates that the polymer solution does infiltrate the interstice among the aligned CNTs. Figure 1f and 1g are the SEM images of the side view of the membrane in a low and high magnification, respectively. The membrane generally has an average thickness of 50 μm with the CNTs (indicated by white arrows) located at the membrane bottom. The pores in the bottom CNT zone are easily visible. Their size is in the range of 50-100 nm, which is very close to that in a CNT-free porous membrane reported previously.[42] The gradient of the degree of electrostatic complexation (DEC) along the cross-section is confirmed by the increasing content of sulfur from top (5.24 wt. %) to the bottom (11.0 wt. %) (Table S1, Supporting Information). As previously reported,[33] the sulfur content exists only in the $Tf_2N^-$ anion and is directly related to the ionic cross-linking density into aqueous solution, as the formation of each ionic crosslinking point is followed by the release of a $Tf_2N^-$ anion, the only sulfur-containing species. Due to the simplicity in this synthetic procedure, the content of PIL-PAA in the composite membrane can be varied in terms of the coating volume of the polymer solution, and the layer number of CNT sheets on the glass plate can be precisely tuned (Figure S5, Supporting Information) to produce membranes with different number of layers of CNTs (Figure S6, Supporting Information).

The resulting membrane bends readily in the transverse direction (90° to the longitudinal axis of the CNTs) with the least bending resistance and the greatest swellability.[43] The as-prepared composite membrane was with a size of 2.5 cm × 1.5 cm bent transversal to CNT alignment direction (Figure S7, Supporting Information). The composite films were readily cut into strips with different CNT orientations but with the same aspect ratio of length/width ~ 11: 3 (**Figure 2**a). When starting from a thin strip film, the membrane exposed to acetone vapor rolls into a tube when the CNTs are oriented along the longitudinal axis of the membrane (Figure 2b), or twists into a ribbon when the CNTs are aligned in a 45° angle (Figure 2c). They will roll into a loop when the CNTs are oriented along the short axis (Figure 2d). The rapid and reversible actuation into targeted geometry towards acetone vapor proves the success of our design concept.

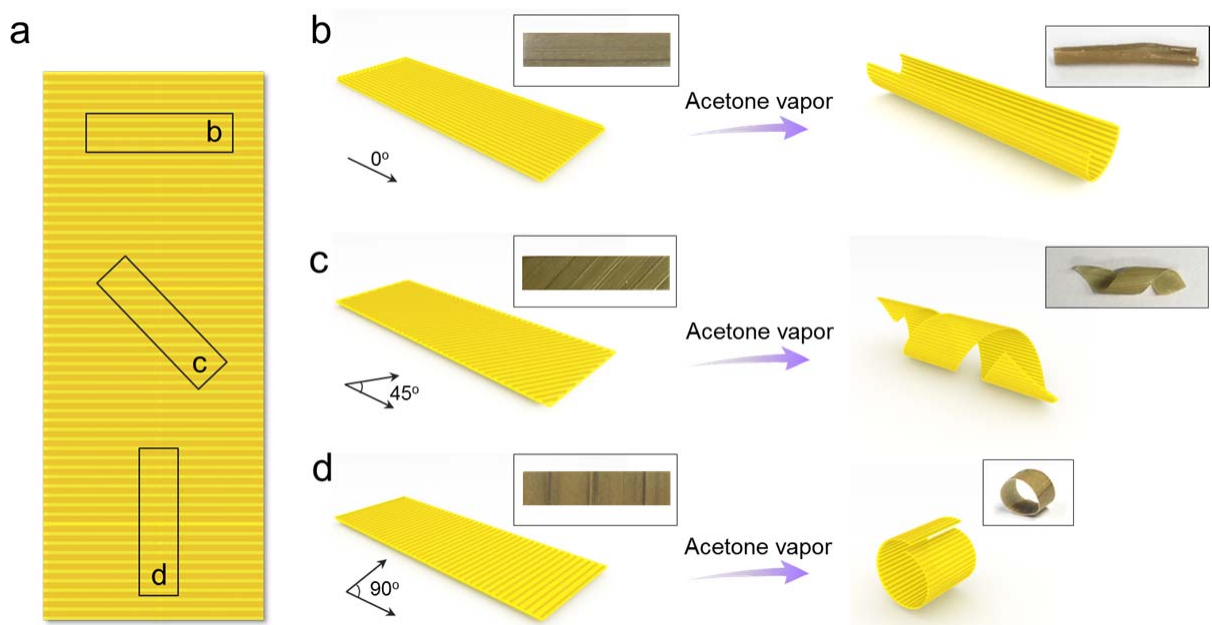

**Figure 2.** Solvent-induced anisotropic bending deformations generated by composite membrane. a) The as-prepared composite membrane with aligned CNTs. CNTs in the membranes are oriented longitudinally (0°) (b), 45° to the longitudinal and transverse direction (c) and transversal (90°) (d) to the length of the membranes. In contact with acetone vapor, the membranes rolled into tubes (b), coiled ribbons (c) and loops (d), respectively.

The reversible actuation was further studied systematically at various well-defined experimental environmets on a composite strip of 20 mm × 2 mm × 50 μm in size and 2 layers of CNTs. As shown in **Figure 3**a and 3b, the composite strip with CNTs transversal to the length of the strip bent into a loop in 6 seconds towards acetone vapor, with a calculated bending curvature of 1.3 mm$^{-1}$. The membrane then recovered its flat state when pulled back to air. If one performs the same experiments at different temperatures the curvatures after 2 seconds are temperature dependent, reaching 0.33, 0.44, 0.57, and 0.85 mm$^{-1}$ in 2 seconds at temperatures of 20, 40, 50, and 60 $^{\circ}$C, respectively (Figure 3c). Furthermore, the change of curvature as a function of solvents including isopropanol, methanol and ethanol other than acetone was also studied. Among these three chosen solvents, isopropanol can sufficiently induce the deformation of the composite strip to achieve a faster bending kinetics than methanol or ethanol (Figure 3d). In addition to these experiments, the dependence of curvature on the addition of acetone into a water bath is further investigated. When the volume ratio of acetone/water was increased, the composite strip rolled into a loop with an increased curvature and a reduced time as shown in Figure 3e. Figure 3f demonstrates that the bending deformation of the composite membrane is reversible for at least 300 cycles with curvature maintained by 80% as well as with the bending direction transversal to the aligned CNTs, but the curvature can be further recovered above 90% after soaking in water for 1 h, a process that washes away the tightly absorbed acetone molecules in the polymer matrix.

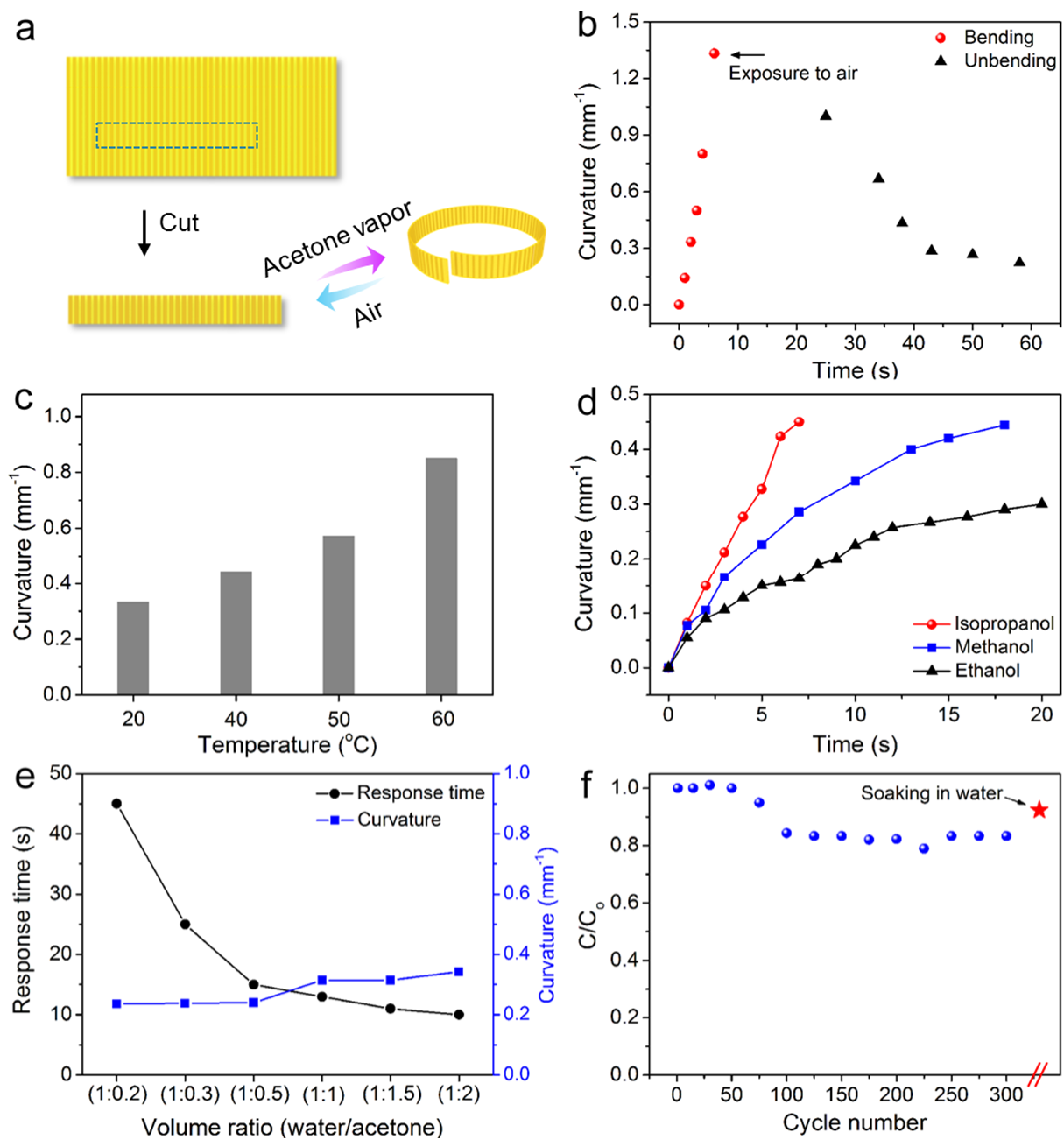

**Figure 3.** a) Actuation behavior of the membrane actuator. b) Plot curvature of the PIL-PAA@CNTs membrane towards acetone vapor and then back to air. c) Curvature of the membrane aftere 2 s towards acetone vapor as a function of temperature. d) Actuation behavior on different solvents. e) Dependence of response time and curvature on the volume ratio of water/acetone. f) Dependence of curvature on the cycle number. The data in star shows the curvature of the membrane was further recovered after soaking in water for 1 h.

Compared with the pristine CNT-free porous PIL membranes, the incorporation of CNTs additionally provided the composite membrane with, besides a controlled actuation direction, a high electric conductivity due to the outstanding electron conductivity of the CNTs. The resistance of the composite membrane was measured both along and across the longitudinal axis of the CNTs (Inset in Table S2, Supporting Information). Along the CNTs the resistance was 0.26 kΩ, whereas in the transverse direction 1.87 kΩ was measured. Remarkably there is a 6-fold difference in the two main directions. Obviously, the composite membrane demonstrates anisotropic electron conduction. Besides, by increasing the layer number of CNT sheets from 1 to 50, the composite membrane shows a rapidly reduced resistance from 301 kΩ to 1.87 kΩ and from 9.93 kΩ to 0.26 kΩ in the transverse and longitudinal directions, respectively (Table S2, Supporting Information). Furthermore, these composite membranes can serve as a current switch in a circuit (Figure S8, Supporting Information). A composite strip with CNTs oriented transversal to the length of the membrane (shown as red dashed line in Figure S8, Supporting Information) displayed an electric conductivity enough to light up a LED lamp. After exposure to acetone vapor, the composite membrane bent away from the fixed copper wire to cut off the circuit because of the actuation movement.

The vapor-induced bending event of the PIL-PAA@CNTs membrane can be further programmed by setting aligned CNTs in various orientations in the composite membrane to design actuation of tunable and complex behaviors. For example, the porous composite membrane was made into an artificial flower to open and close reversibly upon vapor stimulus. The CNTs in porous membranes were designed to be aligned in the direction shown by the red dashed line in Figure S9 (Supporting Information), making only two of the petals to bend towards each other upon exposure to acetone vapor.

It has been reported that aligned CNTs in composite membranes lead to anisotropic mechanical properties such as a higher modulus and strength in longitudinal than in transverse directions.[44-46]. To explore the mechanism of the controlled bending induced by aligned CNTs, samples with both longitudinal and transverse orientation of CNTs were tested in tension (Figure S10a and S10b, Supporting Information). Since the amount of incorporated CNTs is small (2 layers with a rough thickness of 15 nm each), no differences in ultimate tensile stress and only small but nonnegligible differences in tensile stiffness with higher values for samples with longitudinal CNT orientation were found (Figure S10c and S10d, Supporting Information). Despite the small differences in stiffness and the small fraction of CNTs, their presence essentially dictates the actuation style of composite membrane. One possible reason is that as CNTs are located close to the bottom of the membrane, which is far from the centroid/bending plane of the membrane, it would have a proportionally large influence on properties in bending. Furthermore, in a similar way to the role of cellulose microfibrils in plant cell walls,[31] aligned CNTs serve as a geometrical constraint and thus can not only control anisotropy of stiffness[45,46] but also the expansion or swelling of the polymer matrix.[47] It is known that a bilayer undergoing uniform swelling will first bend in all directions uniformly forming a spherical surface.[48] For small swelling strains, the change in Gaussian curvature and bilayer stretching energy is small, however above a critical strain the stretching energy associated with uniform swelling will dominate and the system will bifurcate and bend in only one direction. Such a bifurcation is very sensitive to boundary conditions[48] and it is likely that even small anisotropic properties introduced by the presence of CNTs will cause membrane bending to be favoured in the "soft" direction. By tuning the orientation of the CNTs in the membranes, different actuation behaviors can be achieved. In one example, the aligned CNTs were paved in a membrane with two different orientations such as longitudinal CNTs in the bottom part and the oblique CNTs with 25º angle in the top part, thus leading to bending of the bottom part into a tube and top part from the oblique direction (**Figure 4**a). If the oblique CNTs in the top part of the membrane was replaced with transverse direction (Figure 4b), the top part of the membrane bent from two angles to reach the lowest energy configuration. The concept of the programmable actuation here is similar to plants that can direct shape changes through the controlled orientation of the reinforcing cellulose microfibrils in the plant cells walls. Through this approach, a variety of shape changes can be further engineered by constructing different and complex orientations of CNTs in a single membrane.

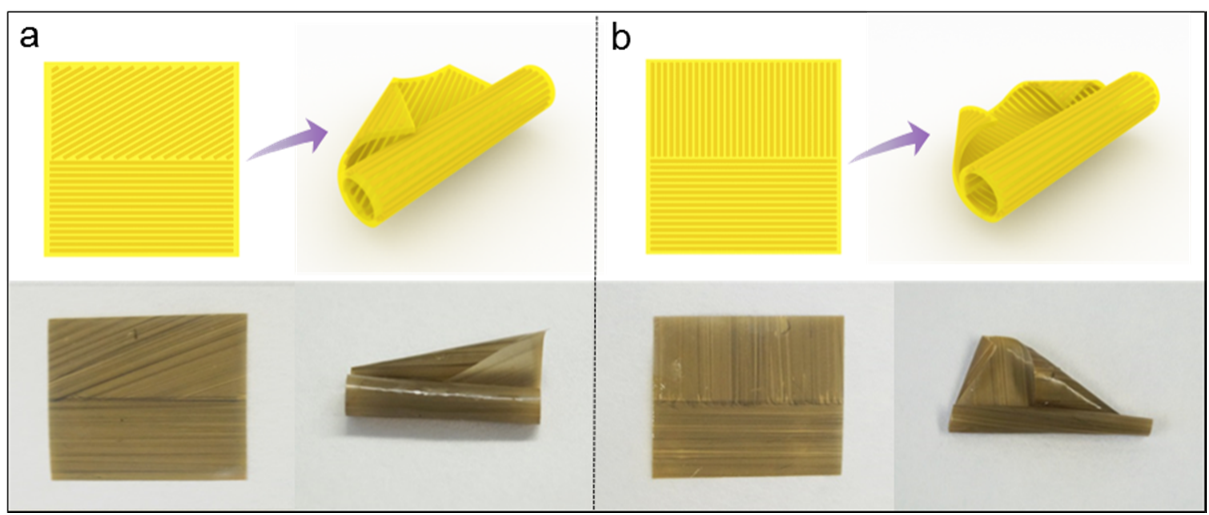

**Figure 4**. Programmable actuation of 2 composite membranes with different CNT orientations.

In summary, a composite porous polymer membrane with aligned CNTs in its matrix is developed which exhibits programmable, anisotropic actuation towards organic vapors. The alignment of CNTs in multiple directions and the gradient porous structure of the membrane are critical for the actuation, and the actuating direction can be well controlled at a direction perpendicular to the longitudinal orientation of CNTs. The reversible actuation may be repeated for 300 cycles without obvious fatigue. These polymer composites are promising building blocks for a wide variety of applications such as sensors and actuators. The concept of programmable actuation succeeded here in a porous polymer mateial *via* complex orientation of CNTs reflects a new high level of control over physical motions that is applicable to a broad range of soft actuators. Nevertheless, it is tough in the present model to quantify the responses of the membrane by the orientation degree of CNTs, the pore distribution and the degree of crosslinking; the thresholds of response time, the critical concentrations of the vapors and the variations of curvature require further optimization for potential sensor application.

**Experimental Section**

*Materials:* Continuous aligned CNT sheets with an apparent density of 1.41 µg/cm$^2$ were dry-drawn from the spinnable CNT arrays synthesized by chemical vapor deposition, a technique which has been previously reported.[41] Lithium bis(trifluoromethane sulfonyl)imide (LiTf$_2$N, 99.95%), Poly(acrylic acid) (PAA, $M_w$ = 1,800 g/mol) and concentrated aqueous ammonia (28 wt. %) were purchased from Sigma-Aldrich and used without further purification. The α,α'-azobisisobutyronitrile (AIBN) was recrystallized from methanol before use.

*Preparation of the PIL-PAA@CNTs membrane:* Typically, the cationic PIL was obtained by free radical polymerization of an IL monomer, 3-cyanomethyl-1-vinylimidazolium bromide, with AIBN as initiator in DMSO at 90 °C for 48 h according to the previous report.[31] The polymerization was then followed by an anion-exchange reaction in aqueous LiTf$_2$N solution. For membrane preparation, the aligned CNT sheets were transferred onto a clean glass plate, and the PIL-PAA solution in DMF (1.0 g of PIL and 0.18 g of PAA dissolved in 10 mL DMF, monomer unit ratio: 1/1) was cast onto the pre-prepared CNT sheets on the glass plate. The glass plate was dried at 80 °C for 3 h and soaked in 0.2 wt. % aqueous ammonia for 2h. The aligned PIL-PAA@CNTs composite membrane was finally removed from the glass plate and cut into different shapes. The actuation experiments were conducted by placing the composite strip with CNT transversal to the length of the membrane above the liquid phase of the solvent; the strip will be bent to a loop towards solvent vapor and then recovered its flat state when pulled back into air. The bending curvature was calculated based on the formula of c = 1/r, r is radius of the bending loop.

*Chemical and structural characterization:* PIL was characterized by $^1$H NMR measurements (Bruker DPX-400 spectrometer), which was carried out using DMSO-d$_6$ as solvent. The PIL had the apparent molecular weight and polydispersity index (PDI) of $1.38 \times 10^5$ g mol$^{-1}$ and 3.34, respectively, determined by gel permeation chromatography (GPC) (eluent: 80% of acetate buffer and 20% of methanol). The structures and morphologies of aligned CNTs and the composite membranes were characterized by using scanning electron microscopy (GEMINI LEO 1550 microscope, 3 kV acceleration voltage) and transmission electron microscopy (TEM, JEOL JEM-2100F operated at 200 kV). The relative sulfur content in the cross-section of the membrane was obtained from the EDX measurements performed on the SEM with an EDX spectrometer. The resistances of the composite membranes were measured with a Benning MM 7-1 multimeter.

*Micromechanical characterization*: 2 mm × 20 mm × 0.05 mm strips with CNTs being aligned in the longitdudinal and transverse direction were cut out of a composite membrane with 2 layers of CNTs. Previous to the test the widths of the strips were measured by light microscopy. Afterwards they were glued with cyanacrylate glue (Loctite 454) on metal sample holders attached with a pin-and-hole assembly (Figure S10a, Supporting Information) to a self built micro-tensile tester equipped with a 2.5 N load cell. Free test length was 15 mm, the pin-and-hole assembly allowed for sample alignment in 2 dimensions. For strain detection by video-extensometry two markers were placed on the sample. The sample area between the two markers was cut with a razor blade after the test and the membrane thickness was determined by scanning eletron microscopy (GEMINI LEO 1550). During the tensile tests the samples were kept in the wet condition. Stresses were calculated based on the cross sectional area of the strips between the two markers.

**Supporting Information**
Supporting Information is available from the Wiley Online Library or from the author.

**Acknowledgements**

This work was supported by the Max Planck Society, the Program of the International Max Planck Research School (IMPRS) and the ERC (European Research Council) Starting Grant with project number 639720 – NAPOLI.

References

[1]     S. Borini, R. White, D. Wei, M. Astley, S. Haque, E. Spigone, N. Harris, J. Kivioja, T. Ryhanen, *ACS Nano* **2013**, *7*, 11166.

[2]     S. H. Hwang, D. Kang, R. S. Ruoff, H. S. Shin, Y. B. Park, *ACS Nano* **2014**, *8*, 6739.

[3]     Z. Jiang, M. Xu, F. Y. Li, Y. L. Yu, *J. Am. Chem. Soc.* **2013**, *135*, 16446.

[4]     P. K. Kundu, G. L. Olsen, V. Kiss, R. Klajn, *Nat. Commun.* **2014**, *5*.

[5]     X. Lu, Z. T. Zhang, H. P. Li, X. M. Sun, H. S. Peng, *J. Mater. Chem. A* **2014**, *2*, 17272.

[6]     Y. J. Men, H. Schlaad, A. Voelkel, J. Y. Yuan, *Polym. Chem.* **2014**, *5*, 3719.

[7]     H. Saito, Y. Takeoka, M. Watanabe, *Chem. Commun.* **2003**, 2126.

[8]     M. A. Yassin, D. Appelhans, R. G. Mendes, M. H. Rummeli, B. Voit, *Chem.-Eur. J.* **2012**, *18*, 12227.

[9]     M. M. Ma, L. Guo, D. G. Anderson, R. Langer, *Science* **2013**, *339*, 186.

[10]    Y. M. Zhang, L. Ionov, *ACS Appl. Mater. Interfaces* **2014**, *6*, 10072.

[11]    F. Zhao, Y. Zhao, H. H. Cheng, L. T. Qu, *Angew. Chem. Int. Ed.* **2015**, *54*, 14951.

[12]    H. Nandivada, A. M. Ross, J. Lahann, *Prog. Polym. Sci.* **2010**, *35*, 141.

[13]    E. Sokolovskaya, S. Rahmani, A. C. Misra, S. Brase, J. Lahann, *ACS Appl. Mater. Interfaces*, **2015**, *7*, 9744.

[14]    M. Fuchiwaki, J. G. Martinez, T. F. Otero, *Adv. Funct. Mater.* **2015**, *25*, 1535.

[15]    B. K. Juluri, A. S. Kumar, Y. Liu, T. Ye, Y. W. Yang, A. H. Flood, L. Fang, J. F. Stoddart, P. S. Weiss, T. J. Huang, *ACS Nano* **2009**, *3*, 291.

[16]    J. J. Liang, Y. Huang, J. Y. Oh, M. Kozlov, D. Sui, S. L. Fang, R. H. Baughman, Y. F. Ma, Y. S. Chen, *Adv. Funct. Mater.* **2011**, *21*, 3778.

[17]    I. S. Romero, N. P. Bradshaw, J. D. Larson, S. Y. Severt, S. J. Roberts, M. L. Schiller,

J. M. Leger, A. R. Murphy, *Adv. Funct. Mater.* **2014**, *24*, 3866.

[18]   N. Terasawa, I. Takeuchi, *Electrochim. Acta* **2014**, *123*, 340.

[19]   X. J. Xie, L. T. Qu, C. Zhou, Y. Li, J. Zhu, H. Bai, G. Q. Shi, L. M. Dai, *ACS Nano* **2010**, *4*, 6050.

[20]   D. Z. Zhou, G. M. Spinks, G. G. Wallace, C. Tiyapiboonchaiya, D. R. MacFarlane, M. Forsyth, J. Z. Sun, *Electrochim. Acta* **2003**, *48*, 2355.

[21]   M. D. Lima, N. Li, M. Jung de Andrade, S. Fang, J. Oh, G. M. Spinks, M. E. Kozlov, C. S. Haines, D. Suh, J. Foroughi, S. J. Kim, Y. Chen, T. Ware, M. K. Shin, L. D. Machado, A.F. Fonseca, J. D. Madden, W. E. Voit, D. S. Galvão, R. H. Baughman, *Science* **2012**, *338*, 928.

[22]   S. Turcaud, L. Guiducci, P. Fratzl, Y. J. M. Bréchet, J. W. C. Dunlop, *Int. J. Mat. Res.* **2011**, *102*, 607.

[23]   Z. L. Wu, M. Moshe, J. Greener, H. Therien-Aubin, Z. Nie, E. Sharon, E. Kumacheva, *Nat. Comm.* **2013**, *4*, 1586.

[24]   G. Stoychev, S. Turcaud, J. W. C. Dunlop, L. Ionov, *Adv. Funct. Mater.* **2013**, *23*, 2295.

[25]   E. Reyssat, L. Mahadevan, *EPL*, **2011**, *93*, 54001.

[26]   H. H. Cheng, J. Liu, Y. Zhao, C. G. Hu, Z. P. Zhang, N. Chen, L. Jiang, L. T. Qu, *Angew. Chem. Int. Ed.* **2013**, *52*, 10482.

[27]   S. Timoshenko, *J. Opt. Soc. Am.* **1925**, *11*, 233.

[28]   L. Guiducci, J. C. Weaver, Y. J. M. Bréchet, P. Fratzl, J. W. C. Dunlop, *Adv. Mater. Interfaces* **2015**, *2*. 1500011.

[29]   P. Fratzl, F. G. Barth, *Nature* **2009**, *462*, 442.

[30]   I. Burgert, M. Eder, N. Gierlinger, P. Fratzl, *Planta*, **2007**, *226*, 981.

[31]   R. M. Erb, J. S. Sander, R. Grisch, A. R. Studart, *Nat. Commun.* **2013**, *4*.

[32]  A. R. Studart, R. M. Erb, *Soft Matter* **2014**, *10*, 1284.

[33]  Q. Zhao, J. W. C. Dunlop, X. L. Qiu, F. H. Huang, Z. B. Zhang, J. Heyda, J. Dzubiella, M. Antonietti, J. Y. Yuan, *Nat. Commun.* **2014**, *5*.

[34]  J. Yuan, D. Mecerreyes, M. Antonietti, *Prog. Polym. Sci.* **2013**, *38*, 1009.

[35]  O. Green, S. Grubjesic, S. Lee, M. A. Firestone, *Polym. Rev.* **2009**, *49*, 339.

[36]  S. Taccola, F. Greco, E. Sinibaldi, A. Mondini, B. Mazzolai, V. Mattoli, *Adv. Mater.* **2015**, *27*, 1668.

[37]  H. Lee, C. G. Xia, N. X. Fang, *Soft Matter* **2010**, *6*, 4342.

[38]  M. Moniruzzaman, K. I. Winey, *Macromolecules* **2006**, *39*, 5194.

[39]  A. B. Dalton, S. Collins, E. Muñoz, J. M. Razal, V. H. Ebron, J. P. Ferraris, J. N. Coleman, B. G. Kim, R. H. Baughman, *Nature* **2003**, *423*, 703.

[40]  J. Y. Yuan, C. Giordano, M. Antonietti, *Chem. Mater.* **2010**, *22*, 5003.

[41]  H. S. Peng, *J. Am. Chem. Soc.* **2008**, *130*, 42.

[42]  Q. Zhao, J. Heyda, J. Dzubiella, K. Täuber, J. W. C. Dunlop, J. Yuan, *Adv. Mater.* **2015**, *27*, 2913.

[43]  T. W. Chou, *Microstructural design of fiber composites,* Cambridge University Press, Cambridge, UK **1992**.

[44]  H. Zhang, L. Qiu, H. Li, Z. Zhang, Z. Yang, H. Peng, *J. Colloid Interface Sci.* **2013**, *395*, 322.

[45]  E. T Thostenson, T.-W. Chou, *J. Phys. D: Appl. Phys.* **2002**, *35*, L77.

[46]  S. Faraji, K. L. Stano, O. Yildiz, A. Li, Y. Zhu, P. D. Bradford, *Nanoscale*, **2015**, *7*, 17038.

[47]  J. Deng, J. Li, P. Chen, X. Fang, X. Sun, Y. Jiang, W. Weng, B. Wang, H. Peng, *J. Am. Chem. Soc.* **2016**, *138*, 225.

[48]  S. Alben, B. Balakrisnan, E. Smela. *Nano Lett.* **2011**, *11*, 2280.

# Supporting Information

**Programmable actuation of porous poly(ionic liquid) membranes by aligned carbon nanotubes**

*Huijuan Lin, Jiang Gong, Michaela Eder, Roman Schuetz, Huisheng Peng, John W. C. Dunlop, and Jiayin Yuan*

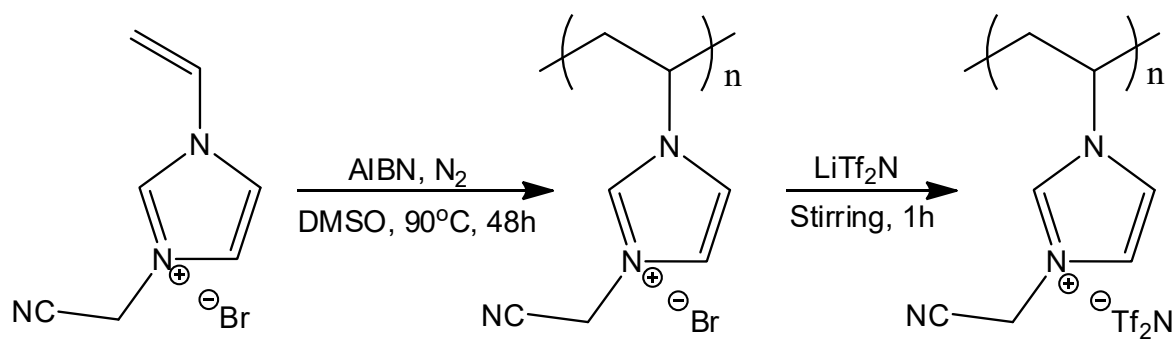

**Figure S1.** Synthesis of the PIL poly[3-cyanomethyl-1-vinylimidazolium bis(trifuoromethane sulfonyl) imide].

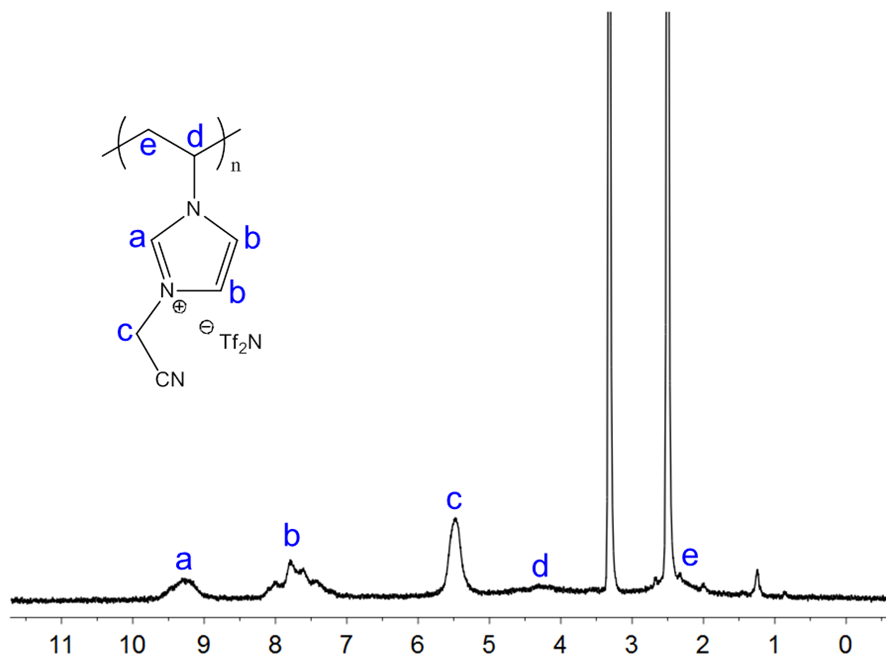

**Figure S2.** $^1$H-NMR spectrum of the used PIL.

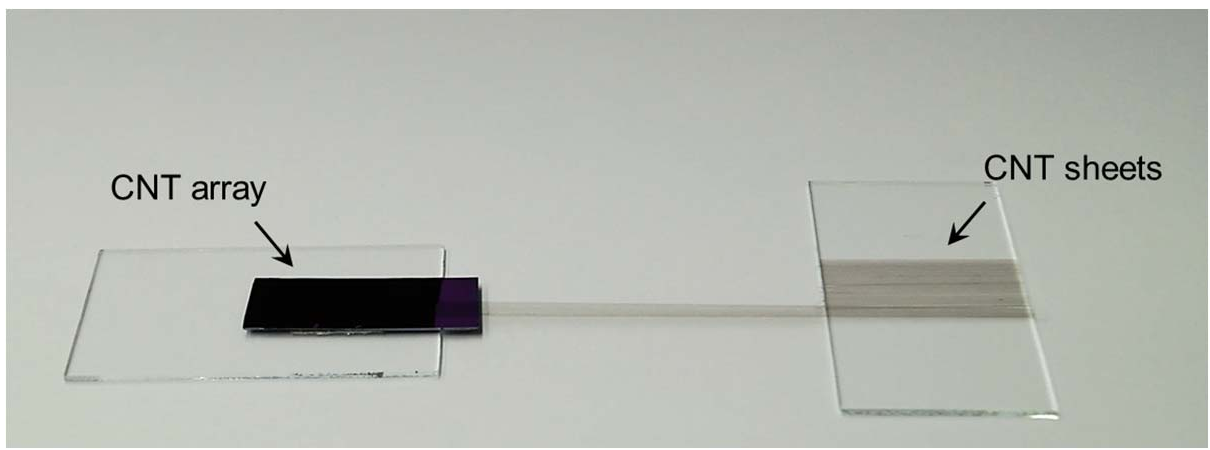

**Figure S3.** Photograph of the experimental setup to prepare the aligned CNT sheets.

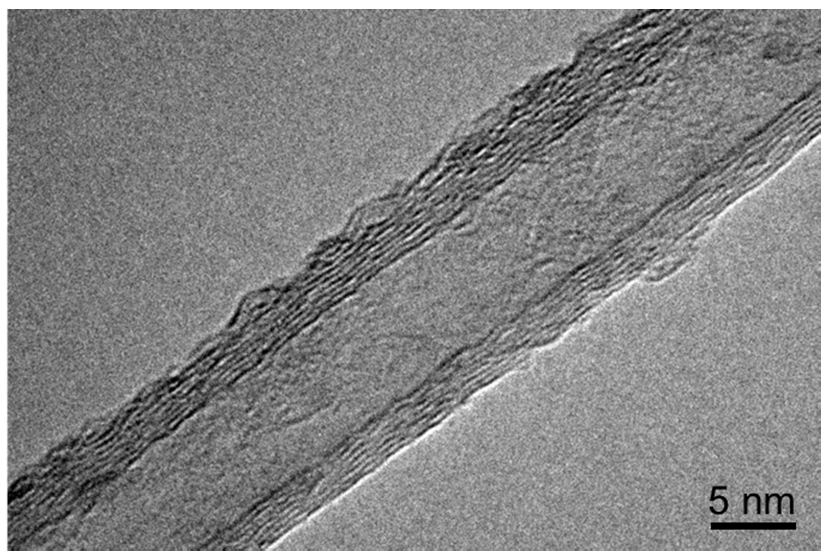

**Figure S4.** High resolution TEM image of a CNT.

| Position | Sulfur Content (wt. %) |
|---|---|
| 1 | 5.24 |
| 2 | 5.93 |
| 3 | 7.11 |
| 4 | 7.80 |
| 5 | 11.0 |

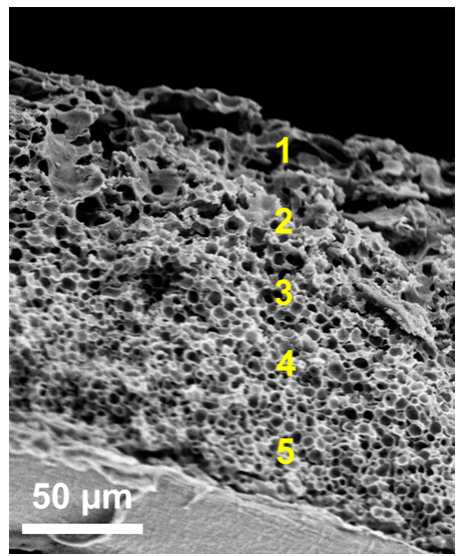

**Table S1**. Sulfur contents from top to bottom in the cross-section from EDX spectra analysis of the PIL-PAA@CNTs porous membrane actuator.

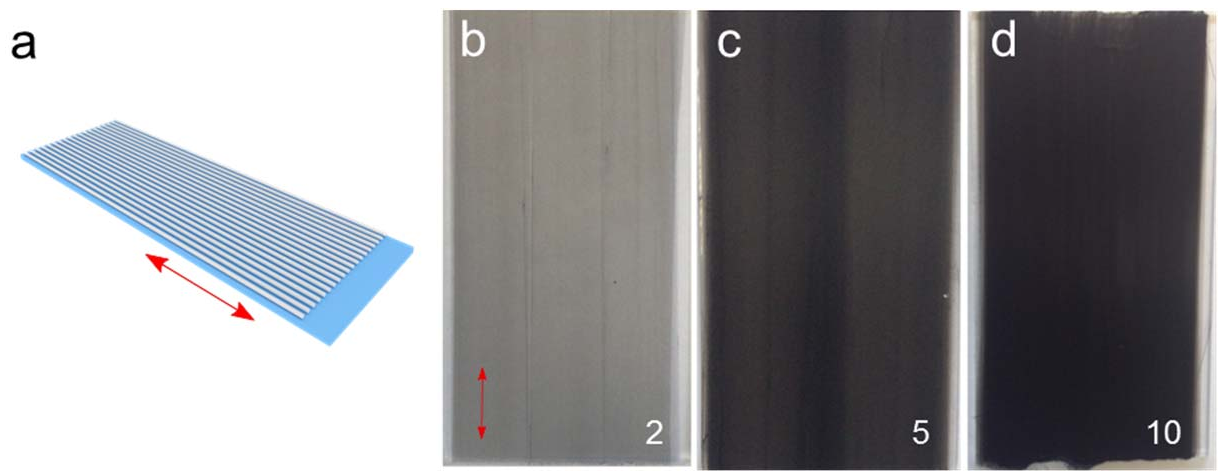

**Figure S5.** a) Schematic illustration of aligned CNT sheets on glass slide. The red arrows represent the longitudinal direction of CNTs. Photographs of aligned CNT sheets: b) 2 layers, c) 5 layers, d) 10 layers.

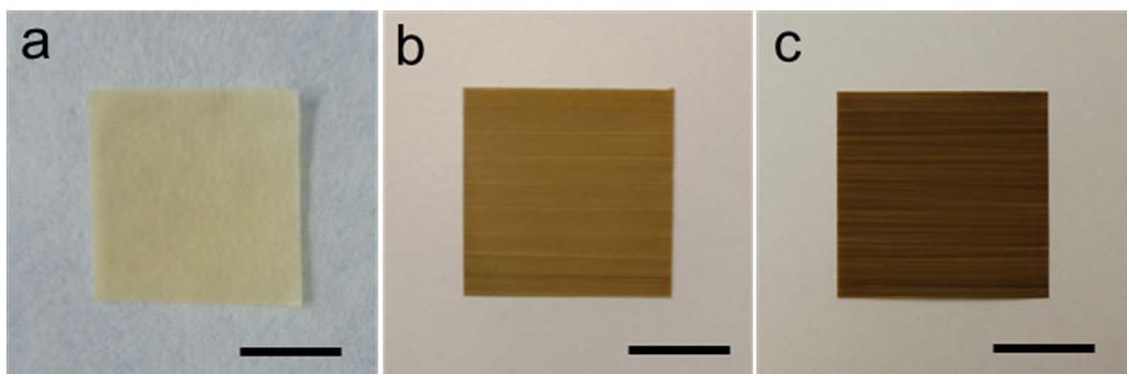

**Figure S6.** Photographs of samples: a) a CNT-free PIL-PAA porous membrane, b) a PIL-PAA@CNTs porous membrane (2 layers of CNTs), c) a PIL-PAA@CNTs porous membrane (5 layers of CNTs). Scar bar: 1 cm.

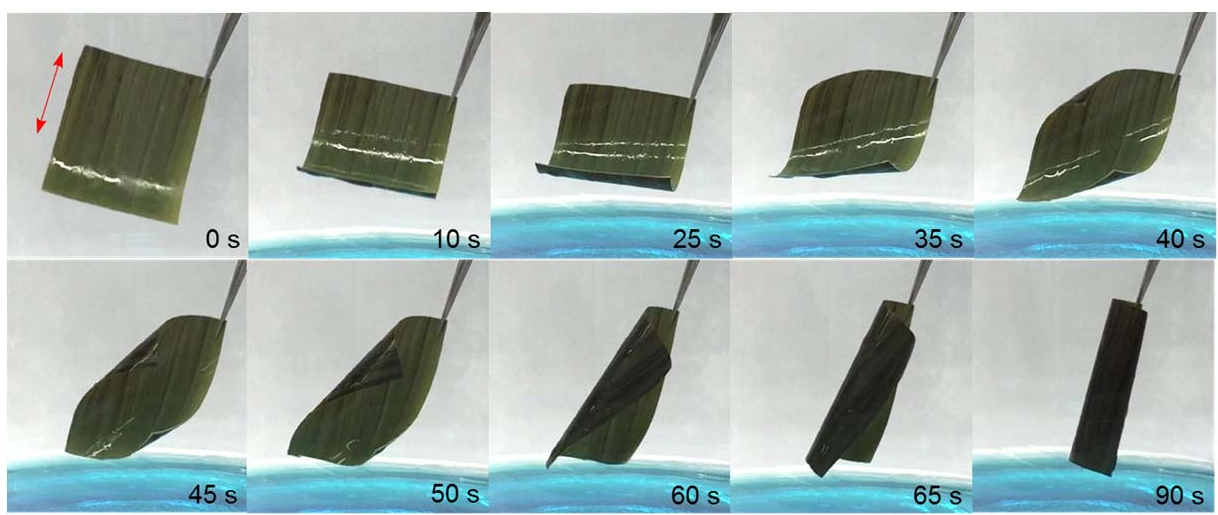

**Figure S7.** The actuation behavior of the as-prepared membrane in acetone vapor. The red arrow represents the longitudinal direction of CNTs.

| Layer | $R_{//}$ (kΩ) | $R_{\perp}$ (kΩ) | $R_{\perp}/R_{//}$ |
|---|---|---|---|
| 1 | 9.93 | 301 | 30.4 |
| 2 | 5.13 | 149 | 29.0 |
| 3 | 2.11 | 43.6 | 20.6 |
| 5 | 1.90 | 15.9 | 14.6 |
| 10 | 0.54 | 6.58 | 12.3 |
| 20 | 0.43 | 3.75 | 8.72 |
| 50 | 0.26 | 1.87 | 7.20 |

**Table S2.** Conductivity of PIL-PAA@CNTs membranes based on different layers of CNT sheets with longitudinal (//) and transversal (⊥) orientions to the CNT alignment direction. Inserted pictures are schematic illustration to the conductivity measurement with silver paste coated on both ends of the membranes as current collector.

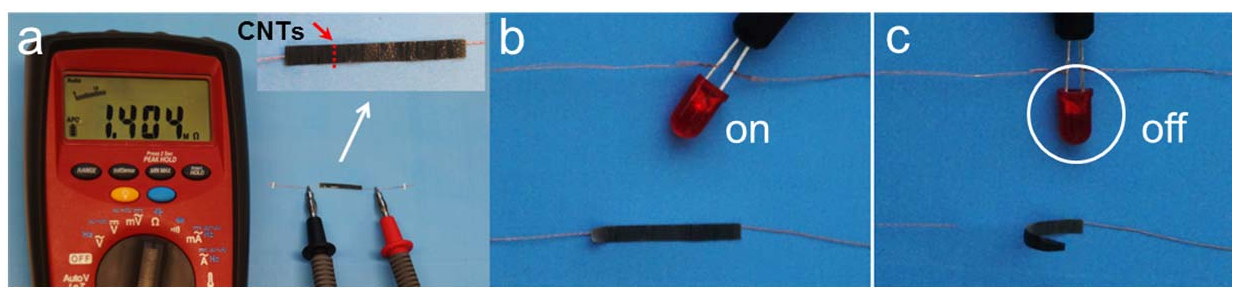

**Figure S8.** a) Conductivity test of the composite membrane. b,c) The conductive composite membrane can be the power switch of the circuit when the membrane was exposed to acetone vapor (24 kPa, 20 °C).

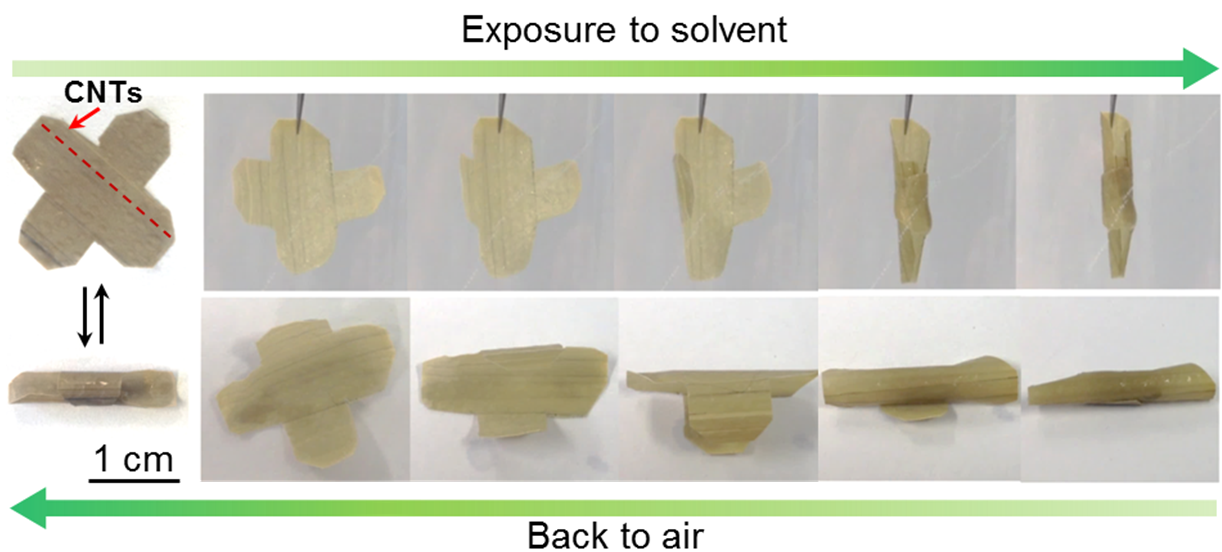

**Figure S9.** Photographs of an artificial flower based on the PIL-PAA@CNTs membranes driven by acetone vapor. The red dashed line represents the longitudinal direction of CNTs.

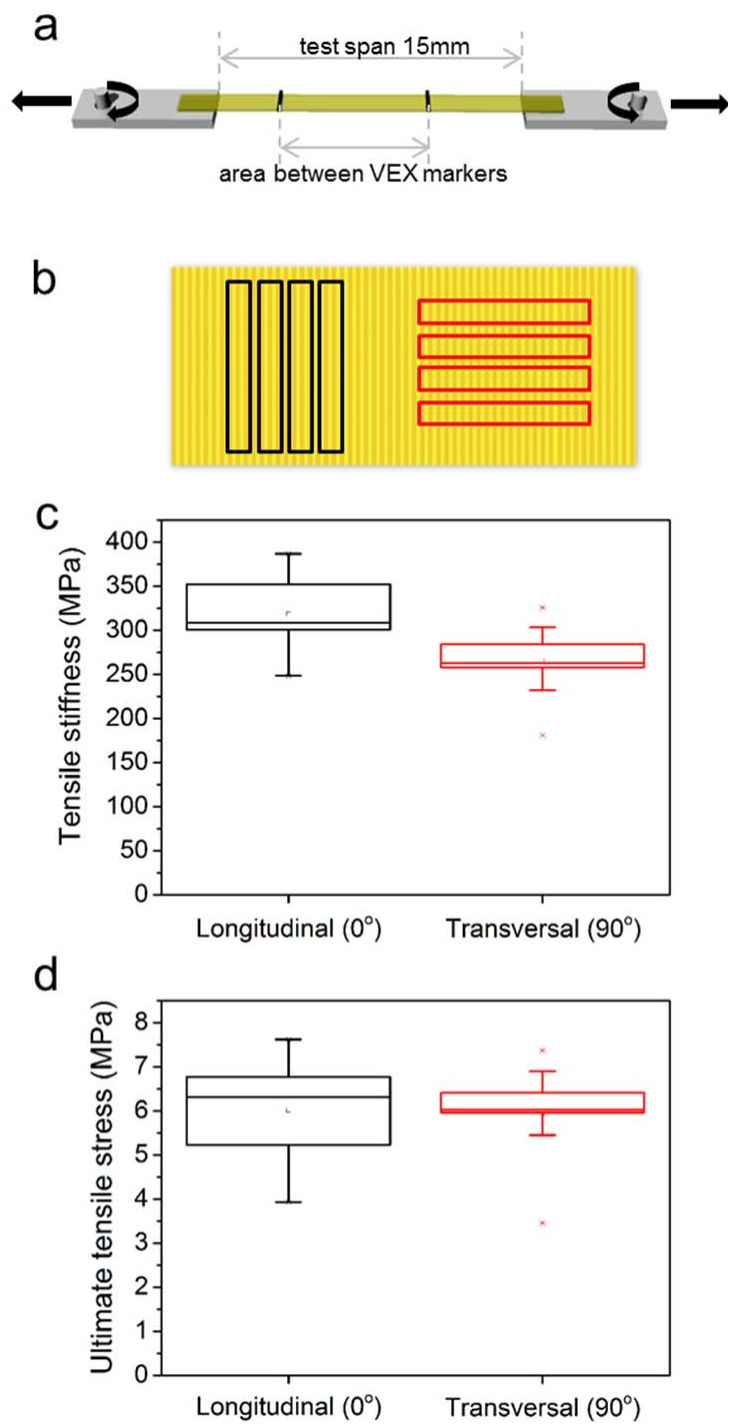

**Figure S10**. a) Schematic drawing of a sample glued on metal holders which were mounted by a pin-hole-assembly during the tensile test. b) Sketch depicting sample preparation out of a composite membrane. c) Tensile stiffness of samples with longitudinally oriented CNTs (black, n = 12) and tranversally oriented CNTs (red, n = 13). d) Ultimate tensile stress of samples with longitudinally oriented CNTs (black, n = 12) and tranversally oriented CNTs (red, n = 13).